\documentclass[a4paper,12pt]{article}

\usepackage{latexsym,amsmath,amsfonts,amssymb}
\usepackage{multirow}
\usepackage[dvips]{graphicx}
\usepackage{mathrsfs,mathtools}
\usepackage{pstricks,pst-node,pst-text,pst-3d}
\usepackage{slashed}
\usepackage{verbatim}
\usepackage{feynmp}
\usepackage[nosort]{cite}
\usepackage[bulletsep]{collref}

\newcommand{\sect}[1]{\setcounter{equation}{0}\section{#1}}

\begin{document}

\begin{titlepage}

\setcounter{page}{0}

\begin{flushright}
{ }
\end{flushright}

\vspace{0.6cm}

\begin{center}
  {\Large \bf Correlators of massive string states with conserved currents}

\vskip 0.99cm

{\bf
George Georgiou$^a$, Bum-Hoon Lee$^b$ and Chanyong Park$^b$
}
\vskip 1.5cm
{\em
${}^a$Demokritos National Research Center, Institute of Nuclear and Particle Physics,
Ag. Paraskevi, GR-15310 Athens, Greece,\\
${}^b$ Center for Quantum Spacetime (CQUeST), Sogang University, Seoul 121-742, Korea
}
\vskip 0.8cm
{\small \sffamily
georgiou@inp.demokritos.gr,\,
bhl@sogang.ac.kr,\,
cyong21@sogang.ac.kr
}

\vskip 1.5cm

\end{center}

\begin{abstract}
We calculate correlation functions of the $R$-current or the stress-energy tensor $T_{\mu\nu}$
with two non-protected operators dual to generic massive string states with rotation in $S^5$, in the context of the AdS/CFT correspondence.
Field theory Ward identities make predictions about the all-loop behaviour of these correlators.
In particular, they restrict the fusion coefficient to be proportional to the R-charge of the operators
or to their dimension, respectively, with certain coefficients of proportionality.
We reproduce these predictions, at strong coupling, using string theory.
Furthermore, we point out that the recently observed strong coupling factorisation of 4-point correlators
is consistent with conformal symmetry and puts constraints on the strong coupling expressions
of 4-point correlators involving R-currents or the stress-energy tensor.
\end{abstract}

\vfill

\end{titlepage}

\sect{Introduction}\label{sec:intro}

The AdS/CFT correspondence~\cite{Maldacena:1998re,Witten:1998qj} claims a duality between
${\cal N}=4$ Super Yang-Mills  (SYM) theory and type-IIB superstring theory on $AdS_5 \times S^5$ background.
Remarkably, both theories appear to be integrable, at least at the planar level.
The presence of integrability allows oneself to hope that one day both theories will be "solved"
and the AdS/CFT correspondence will be proven.
On the field theory side this requires two pieces of information. One needs to be able to identify the conformal dimensions
of all composite operators of the theory, as well as
the structure constants that determine the Operator Product Expansion (OPE) between two primary operators.
Recently, significant progress has been made in the computation of
the planar conformal dimensions of non-protected
operators for any value of the coupling constant,
employing integrability (for a recent review see \cite{Beisert:2010jr}).
On the other hand, much less is known about the structure constants.

Our current knowledge of the structure constants comes from
a perturbative expansion either around $\lambda=0$, or around $\lambda=\infty$ where the IIB string theory is
approximated by a simpler description.  Comparison of the 3-point
correlators among half-BPS operators in these two different limits led
the authors of~\cite{Lee:1998bxa} to conjectured that the corresponding
structure constants are non-renormalised. On the contrary the 3-point
correlators among non-protected operators receive quantum corrections. On the gauge theory side, the authors
of~\cite{Beisert:2002bb,Chu:2002pd,Okuyama:2004bd,Roiban:2004va,Alday:2005nd,Alday:2005kq,Georgiou:2009tp,Georgiou:2012zj,Engelund:2012re}
studied systematically the structure constants and computed the corrections arising from the planar
1-loop Feynman diagrams. In order to evaluate the correction to the structure constants which is of order $\lambda$
one has to take into account
intricate mixing effects related to the fact that the 2-loop eigenstates of the dilatation operator are needed.
The importance of this operator mixing was stressed in \cite{Georgiou:2008vk,Georgiou:2009tp,Georgiou:2011xj}.
Resolution of the mixing made possible a systematic one-loop study  of three-point functions involving
single trace conformal primary operators up to length five \cite{Georgiou:2012zj}.
On the string theory side it is more
difficult to extract information about non-protected OPE coefficients.
This is so because, in the supergravity limit, all non-protected operators acquire
large conformal dimension and decouple. However, one can obtain information about
non-BPS structure constants by studying the BMN limit of type-IIB string theory~\cite{Berenstein:2002jq}.

Recently, another approach to the calculation of n-points correlators
involving non-BPS states was developed
\cite{Yoneya:2006td,Dobashi:2004nm,Tsuji:2006zn,Janik:2010gc,Buchbinder:2010vw,Zarembo:2010rr,Costa:2010rz}.
More precisely, the authors of \cite{Janik:2010gc} argued that it should be possible to
obtain the correlation functions of local operators corresponding to classical
spinning string states, at strong coupling, by evaluating the string action on
a classical solution with appropriate boundary conditions
after convoluting  with the relevant to the classical states wavefunctions.
In \cite{Buchbinder:2010vw,Roiban:2010fe,Ryang:2010bn,Klose:2011rm}, 2-point and 3-point correlators of vertex operators representing
classical string states with large  spins were calculated.
Finally, in a series of papers
\cite{Zarembo:2010rr,Costa:2010rz,Hernandez:2010tg,Russo:2010bt,Georgiou:2010an,Park:2010vs,Bak:2011yy,Lee:2011fe,Bai:2011su,Bissi:2011dc,
Hernandez:2011up,Ahn:2011zg,Bozhilov:2011zp,Arnaudov:2011wq,Ahn:2011dq,Arnaudov:2011ek,Caputa:2012yj,Bozhilov:2012td,Lin:2012ey} the 3-point function coefficients
of a correlator involving a massive string state, its conjugate and a third "light" state
state were computed. This was done by taking advantage of the known classical solutions
corresponding to the 2-point correlators of operators dual to massive string states.

More recently, an intriguing weak/strong coupling match of correlators
involving operators in the $SU(3)$ sector was observed \cite{Escobedo:2011xw}
\footnote{3-point functions were also studied in \cite{Escobedo:2010xs,Escobedo:2011xw,Foda:2011rr,Gromov:2012uv,Bissi:2012vx}
from the perspective of integrability.}. This match was found to hold
for correlators of two non-protected operators in the Frolov-Tseytlin limit and one short BPS operator
\footnote{The authors of \cite{Bissi:2011ha} have calculated the one-loop correction to the structure constants of operators in the SU(2)
sub-sector to find that this agreement is spoiled. One can also employ the BMN limit where agreement at the 1-loop level was found at
\cite{Grignani:2012ur}.}.
In \cite{Georgiou:2011qk}, this weak/strong coupling match was extended to correlators involving operators in the $SL(2,R)$
closed subsector of $N=4$ SYM theory\footnote{For other interesting works on 3-point functions involving operators in the $SL(2)$ sector see
\cite{Plefka:2012rd,Kazakov:2012ar}.}.
Furthermore, by performing Pohlmeyer reduction for classical solutions living in $AdS_2$ but with a prescribed
nonzero energy-momentum tensor the authors of \cite{Janik:2011bd} calculated the AdS contribution to the
three-point coefficient of three heavy states rotating purely in $S^5$
\footnote{The authors of \cite{Buchbinder:2011jr} questioned this result arguing that string solutions with no $AdS_5$ charges
should be point-like in the $AdS_5$ space.}.
In the same spirit, the three-point fusion coefficient of three GKP strings was calculated in \cite{Kazama:2011cp,Kazama:2012is}.
Finally, an interesting approach to holographic three-point functions for operators dual to short string
states was developed in \cite{Minahan:2012fh}.

The plan for the rest of this paper is as follows.
In Section 2, we present the strong coupling calculation of
the 3-point correlator involving an R-current and two scalar operators dual to massive
string states with arbitrary charges in $S^5$.
This particular 3-point correlator obeys a Ward identity which restricts the corresponding structure constant
to be proportional to the R-charge of the non-protected operators to all orders in perturbation theory.
We verify this expectation at the strong coupling regime by employing the AdS/CFT correspondence.
In Section 3, we present the strong coupling calculation of
the 3-point correlator between the stress-energy tensor $T_{\mu\nu}$ and the same two scalar operators of Section 2.
In this case, conformal symmetry requires that the structure constant will be proportional to the conformal dimension of the
massive operators with a certain coefficient of proportionality. As in Section 2, the string theory result agrees perfectly with
the field theory expectation. Both agreements provide non-trivial tests of both
the dictionary of the AdS/CFT correspondence and the AdS/CFT correspondence itself.
Finally, in Section 4 we focus on the 4-point correlators between an R-current or a stress-energy operator and three scalar operators.
We point out that the strong coupling factorisation of \cite{Buchbinder:2010ek}
is consistent with conformal symmetry in the case where one of the light operators is the
R-current or the stress-energy tensor. Furthermore, we analyse the constraints that the aforementioned factorisation imposes on
the  strong coupling form of the above 4-point correlators.

\sect{3-point correlator of an R-current and two massive string states }\label{sec:R-current}
Correlation functions involving the R-symmetry current have been extensively calculated in the context of the AdS/CFT duality.
In fact one of the first checks of the duality was the matching of the parity odd part of the correlator of three R-symmetry operators.
The Adler-Bardeen theorem states that this correlator receives contribution only from one-loop Feynman diagrams.
The very same correlation function was calculated at strong coupling by means of type-IIB supergravity and agreement between the field theory
and supergravity results was found \cite{Freedman:1998tz,Chalmers:1998xr}.
Another class of correlation functions studied was that of an R-symmetry current and two scalar BPS-states \cite{Freedman:1998tz}.
Again the result at strong coupling, as calculated from supergravity, was found to be in perfect agreement with the expectation from field theory.
In this case, the comparison was possible due to the fact that the aforementioned correlator satisfies certain Ward identities
related to the conservation of the R-current.

In all the cases mentioned above one had to restrict himself to BPS operators since only the supergravity approximation was under control.
In this section we evaluate the three-point correlation function of an R-current with two operators dual to semi-classical string states.
The result of the strong coupling calculation is in perfect agreement with the conformal Ward identity providing a check of both
the dictionary and the AdS/CFT correspondence itself.
\subsection{A Ward identity}
It is well-known that conformal symmetry completely fixes the space-time structure of 3-point functions of a vector operator $V_{\mu}$, a scalar operator ${\cal O}_{\Delta}$
and its conjugate ${\bar {\cal O}}_{\Delta}$ as follows \cite{Fradkin:1998}
\begin{eqnarray}\label{vector}
P_{3\mu}=\langle V_{\mu}(x)\,\,{\cal O}_{\Delta}(x_1)\,\,{\bar {\cal O}}_{\Delta}(x_2)\rangle
=\frac{C_{123}(\lambda)}{x_{12}^{2\Delta-2}(x_1-x)^2(x_2-x)^2}E_{\mu}^x(x_1,x_2)
\nonumber \\
E_{\mu}^x(x_1,x_2)=\frac{(x_1-x)_{\mu}}{(x_1-x)^2}-\frac{(x_2-x)_{\mu}}{(x_2-x)^2}.
\end{eqnarray}
$\Delta$ is the conformal dimension of ${\cal O}_{\Delta}$ which has to be an eigenstate of the dilatation operator.
If $V_{\mu}$ is a conserved current, e.g. one of the components of the R-symmetry current $j_{\mu}^R$, and ${\cal O}_{\Delta}$
is also an eigenstate of the symmetry generated by $j_{\mu}^R$ then \eqref{vector} obeys a certain Ward identity that reads
\begin{eqnarray}\label{Ward-R}
\langle\partial^{\mu}j_{\mu}^R(x) \,\,{\cal O}_{\Delta}(x_1)\,\,{\bar {\cal O}}_{\Delta}(x_2) \rangle= \delta^4(x_1-x)
\langle \delta{\cal O}_{\Delta}(x_1)\,\, {\bar {\cal O}}_{\Delta}(x_2)\rangle+\nonumber \\
\delta^4(x_2-x)
\langle {\cal O}_{\Delta}(x_1)\,\, \delta{\bar {\cal O}}_{\Delta}(x_2) \rangle
=J\big(\delta^4(x_1-x)-\delta^4(x_2-x)\big)\langle {\cal O}_{\Delta}(x_1)\, \,{\bar {\cal O}}_{\Delta}(x_2)\rangle.
\end{eqnarray}
In \eqref{Ward-R} $J$ is R-charge of the operator ${\cal O}_{\Delta}$.
By differentiating \eqref{vector} one obtains
\begin{eqnarray}\label{diff-vector}
\partial^{\mu}P_{3\mu}=-C_{123}(\lambda) 2 \pi^2\big(\delta^4(x_1-x)-\delta^4(x_2-x)\big)\frac{1}{x_{12}^{2\Delta}}.
\end{eqnarray}
Comparing \eqref{Ward-R} and \eqref{diff-vector} one gets \footnote{In all equations above we have assumed that ${\cal O}_{\Delta}$ is
normalised to 1.}
\begin{eqnarray}\label{C123-R}
C_{123}(\lambda)=-\frac{J}{2 \pi^2}.
\end{eqnarray}
This provides an all-loop prediction for the fusion coefficient $C_{123}(\lambda)$.
In the next section we will reproduce both the value of \eqref{C123-R} and the space-time structure of the correlator
\eqref{vector} using string theory.

\subsection{ Evaluation of  $\langle j_{\mu}^R(x) \,\,{\cal O}_{\Delta}(x_1)\,\,{\bar {\cal O}}_{\Delta}(x_2) \rangle $  at strong coupling}

In this section we calculate the 3-point function of an R-symmetry current and two operators dual to generic classical string states with rotation
only in $S^5$ by employing the AdS/CFT correspondence. In order to perform the calculation one should identify which supergravity
field is dual to field theory operator
$j_{\mu B}^A,\,\, A,B=1,2,3,4$. Following \cite{Kim:1985ez} we conclude that the dual field is a combination
of the metric with one index in $AdS_5$ and one index in the internal space $S^5$ and the 4-form potential with 3 indices in the internal space
\begin{eqnarray}\label{dual-R}
j_{\mu B}^A \longleftrightarrow (h_{\mu \alpha}, A_{\mu \alpha \beta \gamma}).
\end{eqnarray}
In \eqref{dual-R} $h_{\mu \alpha}$ and $ A_{\mu \alpha \beta \gamma}$ are the fluctuations of these fields around the $AdS_5 \times S^5$ background.
In what follows our string theory computations will be purely at the classical level.
At this level, $A_{\mu\alpha \beta\gamma}$ will not contribute since it couples to fermions and its contribution will be suppressed by $1/\sqrt{\lambda}$
compared to the one coming from the fluctuations of the metric \cite{Zarembo:2010rr}.
One can now expand the metric in spherical harmonics of the sphere
\begin{eqnarray}\label{harmonics}
h_{\mu \alpha}=\sum B_{\mu}^{I_5}(x) Y_{\alpha}^{I_5}(\Omega)=B_{\mu}^{[ij]}(x) Y_{\alpha}^{[ij]}(\Omega)+...,\,\,i,j=1,...,6
\end{eqnarray}
The fact that $j_{\mu B}^A $ transforms in the $\bold 15$ of $SU(4)$ implies that the sugra field dual to the R-current is$B_{\mu}^{[ij]}$, the first term
in the expansion \eqref{harmonics} (see figure 1 of \cite{Kim:1985ez}).
The corresponding spherical harmonic $Y_{\alpha}^{[ij]}(\Omega)$ is just the component of the $[ij]$ Killing vector of $S^5$ along the
$\alpha$ direction\footnote{ Here we enumerate the 15 Killing vectors of $S^5$ by two antisymmetric indices $[ij],\,\,i,j=1,...,6$.}.
Without any loss of generality one can take the component of the R-current which generates rotations in the $[12]$-plane.
Then the corresponding Killing vector is $Y^{[12]}=Y^{[12]\alpha}\frac{\partial}{\partial\phi^{\alpha}}=\frac{\partial}{\partial\phi_1}$.
This is just the Killing vector corresponding to the $\phi_1$-isometry of the
sphere\footnote{We consider a parametrisation of the 5-sphere in terms of the angles $(\gamma,\psi,\phi_1,\phi_2,\phi_3)$ with
$ds^2_{S^5}=d\gamma^2+ \cos^2{\gamma}d\phi_3+\sin^2{\gamma}(d\psi^2+\cos^2{\psi}d\phi_1^2+\sin^2{\psi}d\phi_2^2)$,
where $x_1+i x_2= \sin{\gamma}\cos{\psi}e^{i \phi_1},\,x_3+i x_4= \sin{\gamma}\sin{\psi}e^{i \phi_2},\,x_5+i x_6= \cos{\gamma}e^{i \phi_3}$
with $\sum_{i=1}^6 x_i^2=1$.}.
From the last equation it is obvious that the only non-zero component of $Y^{[12]}$ is $Y^{[12]\phi_1}=1$ which directly gives
$Y^{[12]}_{\phi_1}=g_{\phi_1\phi_1}Y^{[12]\phi_1}=\sin^2{\gamma}\cos^2{\psi}$.
In conclusion we have $Y_{\alpha}^{[12]}=Y^{[12]}_{\phi_1}=\sin^2{\gamma}\cos^2{\psi}$.

We are now in position to evaluate our 3-point correlator. To this end we follow \cite{Zarembo:2010rr} and consider
\begin{eqnarray}\label{exp-val}
\langle B_i(\vec x)  \rangle=\frac{\langle j^{[12]}_i(\vec x) \,\,{\cal O}_{\Delta}(x_1)\,\,{\bar {\cal O}}_{\Delta}(x_2)  \rangle}{\langle{\cal O}_{\Delta}(x_1)\,\,{\bar {\cal O}}_{\Delta}(x_2) \rangle}=\big\langle B_i(\vec x,x_4=0)\frac{1}{{\cal Z}_{string}}\int DX \,\, e^{-S_{string}[X,\Phi]}\big\rangle_{bulk}.
\end{eqnarray}
In \eqref{exp-val} $B_i(\vec x,x_4=0)=B_i^{[12]}(\vec x,x_4=0)$ denotes the boundary value of the sugra field dual to the R-current.
By $\Phi$ we denote all supergravity fields collectively. Also notice that the string action depends not only on the string coordinates $X(\sigma,\tau)$ but on $\Phi$ too.
Because the sugra field $B_i(\vec x,x_4=0)$ is light with respect to the massive string states one can treat the string in first quantised string theory while the sugra field in the supergravity approximation. One can then expand the string action in powers of the sugra fields.
The relevant field for us is $h_{\mu\alpha}$.
\begin{eqnarray}\label{str-exp}
S_{str}=\frac{\sqrt{\lambda}}{4 \pi}\int d^2\sigma\,\,\sqrt{g} g^{\alpha \beta}\,\,\partial_{\alpha}X^M\partial_{\alpha}X^N G_{MN}+(fermions)\nonumber \\
\Rightarrow \frac{\delta S_{str}}{\delta h_{\mu \beta}(X)}=\frac{\sqrt{\lambda}}{2 \pi}\int d^2\sigma\,\,
(\partial_{\tau}X^{\mu}\partial_{\tau}X^{\beta}+\partial_{\sigma}X^{\mu}\partial_{\sigma}X^{\beta})
\end{eqnarray}
Plugging \eqref{str-exp} in \eqref{exp-val} and keeping only the linear in $h_{\mu \beta}$ term we get
\begin{eqnarray}\label{exp-val1}
\langle B_i(\vec x)  \rangle=-\big\langle B_i(\vec x,x_4=0)\frac{\delta S_{str}[X,\Phi=0]}{\delta h_{\mu \beta}(Z)}h_{\mu\beta}(Z)\big\rangle_{bulk}=
\nonumber \\
-\frac{\sqrt{\lambda}}{2 \pi}\int d^2\sigma\,\,
(\partial_{\tau}Z^{\mu}\partial_{\tau}Z^{\beta}+\partial_{\sigma}Z^{\mu}\partial_{\sigma}Z^{\beta})
\big\langle B_i(\vec x,x_4=0) B_{\mu}(z)\big\rangle_{bulk} Y_{\beta}^{[12]}(\Omega),
\end{eqnarray}
where we have substituted the relevant spherical harmonic.
Here we should mention that in the leading approximation the string path integral of \eqref{exp-val} is dominated by the classical solution tunnelling
from boundary to boundary.
In \eqref{exp-val1} one can recognise the bulk-to-boundary propagator for the sugra field\footnote{We should mention that the capital symbols like
$Z$ correspond 10-dinensional coordinates while the small ones like $z$ 5-dimensional coordinates. Finally, symbols with a vector ${\vec x}$
live in the 4-dimensional CFT.}.
This  bulk-to-boundary propagator reads \cite{Freedman:1998tz}
\begin{eqnarray}\label{prop-B}
G_{\mu i}(z;{\vec x},x_4=0)=\frac{\Gamma(4)}{2 \pi^2 \Gamma(2)}
\frac{z_4^2}{z_4^2+({\vec z})^2-{\vec z})^2}\big(\delta_{\mu i}-2
\frac{(z-x)_{\mu} ~ (z-x)_i}{z_4^2+({\vec z}-{\vec x})^2}  \big)
\end{eqnarray}
In \eqref{prop-B} Greek letters $\mu=0,1,2,3,4$ denote one of the AdS directions while Latin ones $i=0,1,2,3$ denotes the directions along the boundary $\partial AdS_5$.
In order to simplify the calculation we take the limit where the BPS operator insertion on the boundary is very far away from the insertions
of the stringy states, that is $x_{i}\rightarrow \infty$. Then the  bulk-to-boundary propagator simplifies to
\[ G_{\mu i}(z;{\vec x},x_4=0) =\frac{3}{\pi^2} \left\{ \begin{array}{ll}
         \frac{z_4^2}{({\vec x}^2)^3} (\delta_{\mu i}-\frac{2 x_{\mu}x_i}{{\vec x}^2})& \mbox{if $\mu=0,1,2,3$};\\
        \frac{z_4^2}{({\vec x}^2)^3}(\frac{-2 z_4(z-x)_i}{{\vec x}^2})\approx 0 & \mbox{if $\mu=4$ since $\delta_{4i}=0$}.\end{array} \right. \]
\begin{itemize}
\item \bf String solution
\end{itemize}
At this point we need to write down the AdS part of the classical string solution propagating from the boundary to the boundary of the AdS space.
Since we want string states which are dual to scalar operators we need to consider only strings with rotation purely in $S^5$.
Consequently the solution is pointlike in AdS. In the  Poincare patch with Euclidean signature it reads \cite{Janik:2010gc}
\begin{eqnarray}\label{str-sol}
z_4=\frac{A}{\cosh(k \tau) },\,\,\,\,z^0=A \tanh(k \tau),\,\,\,\,\gamma,\psi,\phi_1,\phi_2,\phi_3=\gamma,\psi,\phi_1,\phi_2,\phi_3(\sigma,\tau).
\end{eqnarray}
In this solution the spatial separation of the heavy operators living on the boundary is along the Euclidean time direction and reads $|x_{12}^0|=2 A$.
We should stress that our analysis is valid for a generic sting solution with arbitrary charges in $S^5$.

Plugging the expressions for the bulk-to-boundary propagator, the relevant spherical harmonic (see discussion above \eqref{exp-val}) and for the
string solution in \eqref{exp-val1} we obtain after performing the $\tau$ integral
\begin{eqnarray}\label{exp-val2}
\langle B_i(\vec x)  \rangle=-\frac{3\sqrt{\lambda}}{ 2 \pi^3}\frac{1}{({\vec x}^2)^3} (\delta_{\mu i}-\frac{2 x_{\mu}x_i}{{\vec x}^2}) \,\, I^{\mu},\,\,\,\,\nonumber\\
I^{\mu=0}=\int d^2\sigma\,\,
\partial_{\tau}Z^{\mu}\partial_{\tau}Z^{\beta}\,\,Y_{\beta}\,\,z_4^2=\int d^2\sigma\,\,
\partial_{\tau}z^{0}\partial_{\tau}\phi_1\,\,\sin^2{\gamma}\cos^2{\psi}\,\, \frac{A^2}{\cosh^2(k \tau)}=\nonumber\\
\frac{4 A^3}{3}\int d\sigma\,\, \sin^2{\gamma}\cos^2{\psi}\,\,\frac{\partial\phi_1}{\partial \tau}=\frac{4 A^3}{3} J\frac{2 \pi}{\sqrt{\lambda}},
\end{eqnarray}
where in the last equation we have used the expression for the conserved charge related to the $\phi_1$ angle, $J=\frac{\sqrt{\lambda}}{2 \pi}\int d\sigma\,\, \sin^2{\gamma}\cos^2{\psi}\,\,\frac{\partial\phi_1}{\partial \tau}$.
We should mention that because our string solution extends only
in the 0 and 4-directions $\mu$ of \eqref{exp-val2} can be 0 or 4. But as we saw below \eqref{prop-B} in our limit $x_{i} \rightarrow \infty$ the component of the bulk-to-boundary propagator $G_{4i}=0$. Thus the only value $\mu$ can take is 0.
So overall one gets
\begin{eqnarray}\label{exp-val3}
\langle B_i(\vec x)  \rangle=-J\frac{4 A^3}{\pi^2}\frac{1}{({\vec x}^2)^3} (\delta_{0 i}-\frac{2 x_{0}x_i}{{\vec x}^2})
\end{eqnarray}
To be able to compare with field theory one should take the same limit $x_i\rightarrow \infty$ to the corresponding expression \eqref{vector} for the 3-point correlator in field
theory
\footnote{Without taking the $x_i \to \infty$ limit, after some tedious calculations,
we have also reproduced the exact result \eqref{vector} obtained in the CFT. }.
After half page of algebra the leading term is
\begin{eqnarray}\label{vector-lim}
\langle j_{i}^R(x) \,\,{\cal O}_{\Delta}(x_1)\,\,{\bar {\cal O}}_{\Delta}(x_2) \rangle=C_{123}(\lambda) \frac{x_{12}^{\mu}}{x_{12}^{2\Delta-2}}\frac{1}{({\vec x}^2)^3} (\delta_{\mu i}-\frac{2 x_{\mu}x_i}{{\vec x}^2}).
\end{eqnarray}
In order to make contact with \eqref{exp-val3} we have to take into account that the separation of the heavy operators is $|x_{12}^0|=2A$ and divide \eqref{vector-lim} by the 2-point function of the non-BPS operators to obtain
\begin{eqnarray}\label{vector-lim-1}
\frac{\langle j_{i}^R(x) \,\,{\cal O}_{\Delta}(x_1)\,\,{\bar {\cal O}}_{\Delta}(x_2) \rangle}{\langle {\cal O}_{\Delta}(x_1)\,\,{\bar {\cal O}}_{\Delta}(x_2) \rangle}=C_{123}(\lambda) (2A)^3\frac{1}{({\vec x}^2)^3} (\delta_{0 i}-\frac{2 x_{0}x_i}{{\vec x}^2}).
\end{eqnarray}
Direct comparison of \eqref{vector-lim-1} and \eqref{exp-val3} gives the string theory prediction for the fusion coeeficient
\begin{eqnarray}\label{string-C-R}
C_{123}(\lambda>>1)=-\frac{J}{2 \pi^2}.
\end{eqnarray}
This strong coupling result is in perfect agreement with the all-loop prediction based on the field theoretic Ward identity \eqref{C123-R}.
Furthermore, the spacetime structure of the correlator computed by means of AdS/CFT
is the one specified by the conformal invariance of field theory.

\sect{3-point correlator of the stress-energy tensor $ T_{ij}$ and two massive string states }\label{sec:T-current}
In this section we calculate the 3-point function of two operators dual to semi-classical string states and the stress-snergy tensor.
2- and 3-point functions involving the stress-energy tensor have been calculated in the past, particularly in relation
to the conformal anomaly \cite{Howe:1998zi,Arutyunov:1999nw,Erdmenger:1996yc,Osborn:1993cr}.
The form of the the 3-point correlator of  an arbitrary tensor $ V_{ij}$ and two scalar eigenstates of the dilatation operator is
constrained by conformal symmetry up to a scalar coefficient $C_{123}(\lambda)$
\begin{eqnarray}\label{tensor}
\langle V_{ij}(x)\,\,{\cal O}_{\Delta}(x_1)\,\,{\bar {\cal O}}_{\Delta}(x_2)\rangle
=\frac{C_{123}(\lambda)}{x_{12}^{2\Delta-2}(x_1-x)^2(x_2-x)^2} F_{ij}(x,x_1,x_2)
\nonumber \\
F_{ij}(x,x_1,x_2)=E_{i}^x(x_1,x_2)E_{j}^x(x_1,x_2)-\frac{1}{4}\delta_{ij}\frac{x_{12}^2}{(x_1-x)^2(x_2-x)^2},
\end{eqnarray}
where $E_{i}^x(x_1,x_2)$ is a vector defined in \eqref{vector}.
In the case where $V_{ij}=T_{ij}$ conservation of $T_{ij}$ implies that the fusion coefficient is proportional to the
conformal dimension $\Delta(\lambda)$ of the heavy operator
\begin{eqnarray}\label{C123-T}
C_{123}(\lambda)=-\frac{2}{3 \pi^2}\Delta(\lambda).
\end{eqnarray}
This is an all-loop statement which we will verify at strong coupling in the next section.
As in the previous section we will need the leading term of the correlator \eqref{tensor} in the limit $x_i\rightarrow \infty$.
A simple calculation shows that
\begin{eqnarray}\label{tensor-lim}
\langle V_{ij}(x)\,\,{\cal O}_{\Delta}(x_1)\,\,{\bar {\cal O}}_{\Delta}(x_2)\rangle \overset{x_i\rightarrow \infty}{=}
\frac{C_{123}(\lambda)}{x_{12}^{2\Delta-2}}\frac{1}{({\vec x}^2)^4}\big( x_{12}^k x_{12}^l j_{ki}(x)j_{lj}(x)-\frac{1}{4}\delta_{ij} x_{12}^2\big),
\nonumber\\
j_{ij}(x)= \delta_{ij}-2\frac{x_i x_j}{{\vec x}^2}.
\end{eqnarray}
By dividing with the 2-point function and putting our heavy operators along the $0$-direction at a distance of $x^0_{12}=2A$ as we did with our string solution we get
\begin{eqnarray}\label{tensor-lim-1}
\frac{\langle V_{ij}(x)\,\,{\cal O}_{\Delta}(x_1)\,\,{\bar {\cal O}}_{\Delta}(x_2)\rangle}{\langle {\cal O}_{\Delta}(x_1)\,\,{\bar {\cal O}}_{\Delta}(x_2)\rangle} \overset{x_i\rightarrow \infty}{=}
\frac{C_{123}(\lambda)}{({\vec x}^2)^4}(2A)^4\big(  j_{0i}(x)j_{0j}(x)-\frac{1}{4}\delta_{ij} \big).
\end{eqnarray}

\subsection{ Evaluation of  $\langle T_{ij}(x) \,\,{\cal O}_{\Delta}(x_1)\,\,{\bar {\cal O}}_{\Delta}(x_2) \rangle $  at strong coupling}
The calculation for the case in hand proceeds in a similar way to that of section 2.2.
The field dual to the stress-energy tensor $T_{ij}$ is the fluctuations of the metric
$g_{\mu\nu}=g_{\mu\nu}^{AdS}+h_{\mu\nu},\,\,\mu,\nu=0,...,4$ of the AdS spacetime $T_{ij}\longleftrightarrow h_{\mu\nu}$.
It is apparent that in what follows we will need the bulk-to-boundary propagator for the graviton.
This can be read of the solution to the linearised equations of motion in the covariant gauge of de Donder type
$\nabla_{\mu}(h^{\mu}_{\nu}-\frac{1}{2} \delta^{\mu}_{\nu} h)$ \cite{Liu:1998bu,Arutyunov:1999nw}.
\begin{eqnarray}\label{linear-grav}
h^{\mu}_{\nu}(x_4,{\vec x})=k_G \int d^dy K(x,{\vec y})j_{\mu}^i(x-{\vec y})j_{j}^{\nu}(x-{\vec y}){\cal E}_{ij,kl}{\hat h}_{kl}({\vec y}),
\end{eqnarray}
where ${\hat h}_{kl}({\vec y})$ is the value of the graviton on the boundary, $k_G=\frac{d+1}{d-1}\frac{\Gamma(d)}{\pi^{d/2}\Gamma(d/2)}$ and
${\cal E}_{ij,kl}=\frac{1}{2}(\delta_{ik}\delta_{jl}-\delta_{il}\delta_{jk})-\frac{1}{d}\delta_{ij}\delta_{kl}$. Finally, the scalar propagator $K(x,{\vec y})$ and the inversion tensor are given by
\begin{eqnarray}\label{scalar-prop}
K(x,{\vec y})=\frac{x_4^d}{\big(x_4^2+({\vec x}-{\vec y})^2\big)^d},\,\,j_{\mu}^i(x)=\delta_{\mu}^i-2\frac{x_{\mu}x^i}{x^2}.
\end{eqnarray}
In the equations above $d$ is the dimension of the AdS boundary which in our case is $d=4$.

We are now in position to evaluate our 3-point correlator at strong coupling. By following the same steps as in section 2.2 we obtain
\begin{eqnarray}\label{C123-strong}
\langle {\hat h}_{ij}(\vec x)  \rangle=-\big\langle {\hat h}_{ij}(\vec x,x_4=0)\,\,\frac{\delta S_{str}[X,\Phi=0]}{\delta h^{\nu}_{\mu}(Z)}h^{\nu}_{\mu}(Z)\big\rangle_{bulk}=
\nonumber \\
-\frac{\sqrt{\lambda}}{2 \pi}\int d^2\sigma\,\,
(\partial_{\tau}Z^{\mu}\partial_{\tau}Z_{\nu}+\partial_{\sigma}Z^{\mu}\partial_{\sigma}Z_{\nu })\,\,\big\langle {\hat h}_{ij}(\vec x,x_4=0) h^{\nu}_{\mu}(z)\big\rangle_{bulk}.
\end{eqnarray}
Taking into account that the AdS part of the string solution \eqref{str-sol} does not depend on $\sigma$
and keeping only the leading terms in our usual limit
$x_i \rightarrow \infty$
we get
\begin{eqnarray}\label{C123-strong-1}
\langle {\hat h}_{ij}(\vec x)  \rangle=
-k_G\frac{\sqrt{\lambda}}{2 \pi}\int d^2\sigma\,\,\frac{z_4^4}{(\vec x^2)^4}
\partial_{\tau}Z^{\mu}\partial_{\tau}Z_{\nu}\,\,j_{\mu}^k(z-{\vec x})j_{l}^{\nu}(z-{\vec x}){\cal E}_{kl,ij}=\nonumber \\
-k_G \frac{\sqrt{\lambda}}{2 \pi}\frac{1}{(\vec x^2)^4}(\delta_{0}^k-2\frac{x_0x^k}{\vec x^2})(\delta_{l}^0-2\frac{x^0x_l}{\vec x^2})
{\cal E}_{kl,ij}\,\,
\int d^2\sigma\,\,z_4^4 \partial_{\tau}z^{0}\partial_{\tau}z_{0}
\end{eqnarray}
In order to proceed we need to calculate the integral and simplify the tensorial structure in \eqref{C123-strong-1}.
This is easily done
\begin{eqnarray}\label{int-T}
\int d^2\sigma\,\,z_4^4 \partial_{\tau}z^{0}\partial_{\tau}z_{0}=\int d^2\sigma\,\,z_4^2 (\partial_{\tau}z^{0})^2=A^4\int d\sigma d\tau \frac{k^2}{\cosh^k\tau}=A^4 2\pi k \frac{16}{15}\nonumber\\
(\delta_{0}^k-2\frac{x_0x^k}{\vec x^2})(\delta_{l}^0-2\frac{x^0x_l}{\vec x^2})
{\cal E}_{kl,ij}=(\delta_{0i}-2\frac{x_0 x_i}{\vec x^2})(\delta_{j}^0-2\frac{x^0x_j}{\vec x^2})-\frac{1}{4} \delta_{ij}
\end{eqnarray}
In \eqref{int-T} $i,j=0,1,2,3$ since we are interested in the components of ${\hat h}$ which are parallel to the boundary.
Inserting \eqref{int-T} in \eqref{C123-strong-1} one obtains
\begin{eqnarray}\label{C123-strong-2}
\langle {\hat h}_{ij}(\vec x)  \rangle=-k_G\sqrt{\lambda}\frac{1}{({\vec x^2})^4} A^4 \frac{16}{15} k \big( (\delta_{0i}-2\frac{x_0 x_i}{\vec x^2})(\delta_{j}^0-2\frac{x^0x_j}{\vec x^2})-\frac{1}{4} \delta_{ij}\big)
\end{eqnarray}
Direct comparison of \eqref{C123-strong-2} and \eqref{tensor-lim-1}
shows that they have the same spacetime structure\footnote{Similarly to the vector operator,
\eqref{tensor} can be  exactly reproduced in string calculation
without taking the $x_i \to \infty$ limit.} and that the strong coupling value of the fusion coefficient is
\begin{eqnarray}\label{C123-final}
C_{123}(\lambda>>1)=-\frac{5\Gamma(4)}{3 \pi^2}\frac{1}{15}\sqrt{\lambda}k=-\frac{2}{3\pi^2}\sqrt{\lambda}k=-\frac{2}{3\pi^2}E,
\end{eqnarray}
since $E=\sqrt{\lambda}k$ is the energy of the string \eqref{str-sol} with arbitrary charges in $S^5$.
Thus the string theory result \eqref{C123-final} is in complete agreement with the all-loop field theory expectation \eqref{C123-T} based on Ward
identities since according to the AdS/CFT correspondence $\Delta(\lambda)=E(\lambda)$.

An important comment is in order. According to the AdS/CFT correspondence,
the field theory correlators we are considering, $\langle j_{\mu}^R \,\,{\cal O}_{\Delta}\,\,{\bar {\cal O}}_{\Delta} \rangle$  and
$\langle T_{\mu\nu} \,\,{\cal O}_{\Delta}\,\,{\bar {\cal O}}_{\Delta} \rangle $, should be equal to correlators of the corresponding
string vertex operators which schematically are of the form $\langle V_{\mu}\,\, V_{heavy}\,\,V_{heavy}\rangle$ and
$\langle V_{\mu\nu}\,\, V_{heavy}\,\,V_{heavy} \rangle$ \cite{Buchbinder:2010vw,Buchbinder:2010ek}.
Here $V_{\mu}$ and ${V_{\mu\nu}}$ are the vertex operators for the massless string modes that
correspond to $j_{\mu}$ and $T_{\mu\nu}$ respectively, while $V_{heavy}$ are the vertex operators for the massive string states
which are dual to  primary operators in field theory. The semiclassical value of the aforementioned string correlators
takes the form of a ''light'' vertex operator integrated over the classical string solution sourced by $V_{heavy}$
(see \eqref{exp-val1} and \eqref{C123-strong}) \cite{Buchbinder:2010ek}.
One could have determined these ''light'' vertex operators explicitly.
Then integrating the ''light'' vertex operators over the classical string solution would have given results identical to ours.
In fact, one should understand the analysis following \eqref{exp-val} and \eqref{C123-strong}
as an effective way to determine the contribution coming from these supergravity vertex operators.

\sect{4-point correlators involving the stress-energy tensor or the R-current at strong coupling}\label{sec:TR-currents}
Recently, it was observed that the 4-point function of two operators dual to massive string states and two
light\footnote{By light we mean operators with conformal dimensions $\Delta<<\sqrt{\lambda}$.} BPS operators dual to sugra states
factorises as a product of two 3-point correlators for very large values of the coupling constant $\lambda>>1$.
More precisely, in \cite{Buchbinder:2010ek} it was argued that
\begin{eqnarray}\label{4-point}
&&\langle {\cal O}_{L1}(x_3) {\cal O}_{L2}(x_4)\,\,{\cal O}_{\Delta}(x_1)\,\,{\bar {\cal O}}_{\Delta}(x_2)\rangle=\nonumber\\
&&\frac{\langle {\cal O}_{L2}(x_4)\,\,{\cal O}_{\Delta}(x_1)\,\,{\bar {\cal O}}_{\Delta}(x_2)\rangle \langle {\cal O}_{L1}(x_3)\,\,{\cal O}_{\Delta}(x_1)\,\,{\bar {\cal O}}_{\Delta}(x_2)\rangle}{\langle{\cal O}_{\Delta}(x_1)\,\,{\bar {\cal O}}_{\Delta}(x_2)\rangle}
+O(\sqrt{\lambda}).
\end{eqnarray}
One can imagine that in the place of the massive string states one takes two BPS operators with a very large angular momentum in $S^5$
such that these operators can have a description in terms of a classical point like string state. One can then use \eqref{4-point}
to calculate the strong coupling limit of the 4-point correlator.
The result obtained this way disagrees with the result calculated in the supergravity approximation \cite{Uruchurtu:2008kp} when the charges of the two
''heavy'' BPS states are extrapolated to very large values.
This is so because the supergravity approximation is not a priori valid if the charges
of the supergravity states scale as $\sqrt{\lambda}$. Thus correlators
involving ''heavy'' string modes  cannot be compared to supergravity correlators.
It might happen that in particular cases one can extrapolate the supergravity result to
very large values of the charges ($J \sim \sqrt{\lambda}$) but
that should not be expected in general\footnote{We thank Arkady Tseytlin for bringing this point to our attention.}.

In this section we point out that this strong coupling factorisation
of 4-point correlators is consistent with conformal symmetry in the case where
one of the light operators is the R-current or the stress-energy tensor\footnote{For weak/strong matching of 4-point correlation functions see \cite{Caetano:2011eb}.}.
Furthermore, we will analyse the constraints
on the form of 4-point correlators that are imposed from the aforementioned factorisation at strong coupling.
We start with the 4-point  correlation function among a conserved current $J_{\mu}$ and three scalar operators one of which ${\cal O}_{\delta}$
is not charged under the symmetry generated by $J_{\mu}$ while the other two ${\cal O}_{\Delta}$ and ${\bar {\cal O}}_{\Delta}$ have charges $J$ and $-J$ respectively. In that case the 4-point correlator can be expressed
 as follows \cite{Fradkin:1978}
\begin{eqnarray}\label{4-point-R}
&& \langle {\cal O}_{\Delta}(x_1)\,\,{\bar {\cal O}}_{\Delta}(x_2) {\cal O}_{\delta}(x_3) j_{\mu}(x_4)\rangle= \\
&&\Big( \xi^{h-1}K_{\mu}^{x_4}(x_1,x_3)F(\xi,\eta;\lambda)- \eta^{h-1}K_{\mu}^{x_4}(x_2,x_3)F(\eta,\xi;\lambda)\Big) \langle{\cal O}_{\Delta}(x_1)\,\,{\bar {\cal O}}_{\Delta}(x_2) {\cal O}_{\delta}(x_3)\rangle\nonumber,
\end{eqnarray}
where $h=D/2$ and $D$ is the dimension of the spacetime while $\xi$ and $\eta$ are the conformal cross-ratios and $K_{\mu}$ is given by
\begin{eqnarray}\label{cross-ratios}
\xi=\frac{x_{12}^2 x_{34}^2}{x_{13}^2 x_{24}^2},\,\, \eta=\frac{x_{12}^2 x_{34}^2}{x_{14}^2 x_{23}^2}\nonumber \\
K_{\mu}^{x_4}(x_1,x_3)=\frac{\Gamma(h) J}{2 \pi^h}\frac{x_{13}^2}{x_{14}^2x_{34}^2}(\frac{(x_{14})_{\mu}}{x_{14}^2}-\frac{(x_{34})_{\mu}}{x_{34}^2}).
\end{eqnarray}
Note that conformal symmetry is not enough to fully constrain the spacetime structure of the 4-point correlator
as was the case with the 3-point function. Instead there is an arbitrary function of the cross-ratios $F(\eta,\xi;\lambda)$
which has to be determined by direct calculation.

However, at strong coupling the factorisation \eqref{4-point} implies that \eqref{4-point-R} leads to
\begin{eqnarray}\label{fact}
\xi^{h-1} K_{\mu}^{x_4}(x_1,x_3)F(\xi,\eta;\lambda)-\eta^{h-1} K_{\mu}^{x_4}(x_2,x_3)F(\eta,\xi;\lambda)=\langle{\cal O}_{\Delta}(x_1)\,\,{\bar {\cal O}}_{\Delta}(x_2) j_{\mu}(x_4)\rangle x_{12}^{2\Delta}
\end{eqnarray}
This equation can be satisfied only if its left hand side is independent of $x_3$. This can only happen if
\begin{eqnarray}\label{condition}
\xi^{h-1} \frac{x_{13}^2}{x_{14}^2x_{34}^2}F(\xi,\eta;\lambda)=\eta^{h-1} \frac{x_{23}^2}{x_{24}^2x_{34}^2}F(\eta,\xi;\lambda)
\end{eqnarray}
because in such case $x_3$ disapperas from the parenthesis of \eqref{4-point-R}.
Plugging the value $h=2$ in \eqref{condition} we obtain
\begin{eqnarray}\label{condition-1}
F(\xi,\eta;\lambda)=F(\eta,\xi;\lambda),\,\, {\rm for} \,\,\lambda>>1.
\end{eqnarray}
Thus we conclude that at strong coupling the function $F(\xi,\eta;\lambda)$ should be symmetric under the exchange $\xi \leftrightarrow \eta$.
In the case where $j_{\mu}$ is  the R-current direct comparison of \eqref{condition-1} and \eqref{vector}, \eqref{C123-R} leads to
the conclusion that $F$ is not only symmetric but also completely independent of the cross-ratios $\xi,\eta$.
\begin{eqnarray}\label{final-F}
F(\xi,\eta;\lambda)=F(\eta,\xi;\lambda)=-\frac{J}{2\pi^2}+ O(\xi,\eta;(\sqrt{\lambda})^0).
\end{eqnarray}

We now turn to the last correlation function to be considered. This is the similar to \eqref{4-point-R} with the stress-energy tensor in the place
of the R-current. As above conformal symmetry constrains to some extent this 4-point function \cite{Fradkin:1978}
\begin{eqnarray}\label{final-F}
\langle{\cal O}_{\Delta}(x_1)\,\,{\bar {\cal O}}_{\Delta}(x_2) {\cal O}_{\delta}(x_3) T_{\mu\nu}(x_4)\rangle=
\Big( K_{\mu\nu}^{x_4}(x_1,x_2){\hat F}(\xi,\eta;\lambda)+ K_{\mu\nu}^{x_4}(x_1,x_3)\phi(\xi,\eta;\lambda)+\nonumber \\
K_{\mu\nu}^{x_4}(x_2,x_3)\phi(\eta,\xi;\lambda)\Big) \langle{\cal O}_{\Delta}(x_1)\,\,{\bar {\cal O}}_{\Delta}(x_2) {\cal O}_{\delta}(x_3)\rangle,
\end{eqnarray}
where
\begin{eqnarray}\label{Kmn}
K_{\mu\nu}^{x_4}(x_1,x_2)=\frac{x_{12}^2}{x_{14}^2 x_{24}^2}F_{\mu\nu}(x_4,x_1,x_2),
\end{eqnarray}
where $F_{\mu\nu}$ is defined in \eqref{tensor}.
Strong coupling factorisation results to
\begin{eqnarray}\label{fact-T}
K_{\mu\nu}^{x_4}(x_1,x_2){\hat F}(\xi,\eta;\lambda)+ K_{\mu\nu}^{x_4}(x_1,x_3)\phi(\xi,\eta;\lambda)+
K_{\mu\nu}^{x_4}(x_2,x_3)\phi(\eta,\xi;\lambda)=\nonumber\\
\langle{\cal O}_{\Delta}(x_1)\,\,{\bar {\cal O}}_{\Delta}(x_2) T_{\mu\nu}(x_4)\rangle x_{12}^{2\Delta}.
\end{eqnarray}
Comparing the last equation to \eqref{tensor}and \eqref{C123-T} we conclude that
\begin{eqnarray}\label{fact-T-result}
\phi(\xi,\eta;\lambda)=\phi(\eta,\xi;\lambda)=0\sqrt{\lambda}+O(\xi,\eta;(\sqrt{\lambda})^0)\nonumber\\
{\hat F}(\xi,\eta;\lambda)=-\frac{2\Delta(\lambda>>1)}{3\pi^2}+O(\xi,\eta;(\sqrt{\lambda})^0).
\end{eqnarray}
As in the case of the R-current correlators the leading in large $\lambda$ expressions for the undetermined by the conformal symmetry functions
${\hat F}(\xi,\eta;\lambda), \,\,\phi(\xi,\eta;\lambda)$ are independent from the cross-ratios $\xi$ and $\eta$. We do not expect that this property will hold when $\alpha'$ corrections are taken into account or at the weak coupling expansion.


\vspace{1cm}

\noindent {\large {\bf Acknowledgments}}

\vspace{3mm}

\noindent
We wish to thank George Savvidy and Dimitrios Zoakos and especially Arkady Tseytlin for useful discussions and comments.
This work was partly supported by the
General Secretariat for Research and Technology of Greece and from the European Regional Development Fund (NSRF 2007-13 ACTION, $KPH\Pi I \Sigma$).

This work was also partly supported by the National Research Foundation of Korea(NRF) grant funded by
the Korea government(MEST) through the Center for Quantum Spacetime(CQUeST) of Sogang
University with grant number 2005-0049409 and Basic Science Research Program through the
National Research Foundation of Korea(NRF) funded by the Ministry of
Education, Science and Technology(2010-0022369).



\bibliographystyle{nb}
\bibliography{botany}

\end{document}